# Transport mechanism in amorphous molybdenum silicide thin films

Zhengyuan Liu, Bingcheng Luo [a)], Junbiao Hu, Cheng Xing

School of Physical Science and Technology, Northwestern Polytechnical University, Xi'an, Shaanxi, 710072, China

[a)] Electronic mail: luobingcheng@nwpu.edu.cn

**Abstract**

Amorphous molybdenum silicide compounds have attracted significant interest for potential device applications, particularly in single-photon detector. In this work, the temperature-dependent resistance and magneto-resistance behaviors were measured to reveal the charge transport mechanism, which is of great importance for applications but is still insufficient. It is found that Mott variable hopping conductivity dominates the transport of sputtered amorphous molybdenum silicide thin films. Additionally, the observed magneto-resistance crossover from negative to positive is ascribed to the interference enhancement and the shrinkage of electron wave function, both of which vary the probability of hopping between localized sites.

**Introduction**

Electronic transport in solids is an old but interesting subject, which is generally determined by external and internal factors of the material. Particularly, the existence of lattice periodicity, as an internal factor, would determine the state of the electrons inside the material. For example, the electron wave function shows an extended state when the periodicity exists, and it becomes a localized state when the periodicity is missing [1]. Meanwhile, external stimulus, such as temperature, illumination, and magnetic field, could also change the movement of electrons [2-5]. Due to the absence of lattice periodicity, amorphous materials are naturally chosen as a prototypical platform to study how the external stimulus affects the electronic transport in the localized state [6-8]. Amorphous molybdenum silicide ($a$-MoSi), a typical transition-metal-based amorphous semiconductor, has shown great prospects in optoelectronic devices recently, especially in single-photon detectors [9-12]. Owing to the unique and excellent properties including the tunable superconducting transition temperature [10], intrinsically low flux pinning and homogeneity [11,13], $a$-MoSi based devices surpass both crystalline and conventional amorphous counterparts in detection efficiency, response time and timing jitter [14-16]. Additionally, the attractiveness of $a$-MoSi in vortex dynamics [17] proves the tantalizing research interest.

So far, considerable efforts have been devoted to studying the properties towards the detector application of $a$-MoSi. Specifically, it mainly focuses on realizing the modification of superconducting transition temperature, thereby improving the performance of the device [9,10,12,14,15]. By contrast, investigation on electronic transport under the external field (to a larger extent) is still insufficient, but it is important for developing a comprehensive physical insight into transport behaviors and further exploring their potential applications [9]. Herein, we studied the electronic transport mechanism of sputtered *a*-MoSi thin films through temperature-dependent resistance and magnetoresistance (MR) measurements. One may notice that MR measurement was usually considered as a powerful method to reveal the underlying transport mechanism, because the electron wave function, especially the localized electron wave function, can be altered by the applied magnetic field [5]. Although MR investigation in amorphous compounds would be traced back to 1970s [18-20], covering many aspects like conventional MR, colossal MR and giant MR, the report of analogous MR behavior is still lacking in *a*-MoSi, which is the additional motivation of this work.

**Experimental**

Amorphous molybdenum silicide films with thickness of ~30 nm were deposited on commercial silicon ($>10^5 \, \Omega \cdot cm$) substrates at room temperature and without extra substrate heating by radio-frequency (RF) magnetron sputtering technique. The base pressure of the chamber is less than $1\times10^{-4}$ Pa. The sputtering pressure is 20 mTorr in pure argon ambient. Prior to deposition, the target is pre-sputtered for 30 min to eliminate the influence of target surface. To study the transport mechanism in the localized state system with expected stoichiometry, the RF power density could not exceed 2 W · cm$^{-2}$. A series of samples prepared under different sputtering power exhibited similar resistance-temperature behavior and thus the typical sample prepared under sputtering power density of 2 W·cm$^{-2}$ was chosen for transport investigation in the present case. A commercial atomic force microscopy (AFM) (MFP-3D, Asylum Research) was used to measure the surface topography. The structure and chemical composition were determined by grazing incidence X-ray diffraction (GIXRD, PAN alytical Empyrean) and X-ray photoelectron spectroscopy (XPS, ESCALAB 250), respectively. The temperature-dependent resistances were measured by a physical property measurement system (Cryogenic CFMS-14T) using the four-probe method, with the temperature regime from liquid nitrogen to room temperature. Meanwhile, the perpendicular magnetic field was applied, and the MR measurement was done at selected temperature in the same measurement system.

**Results and Discussion**

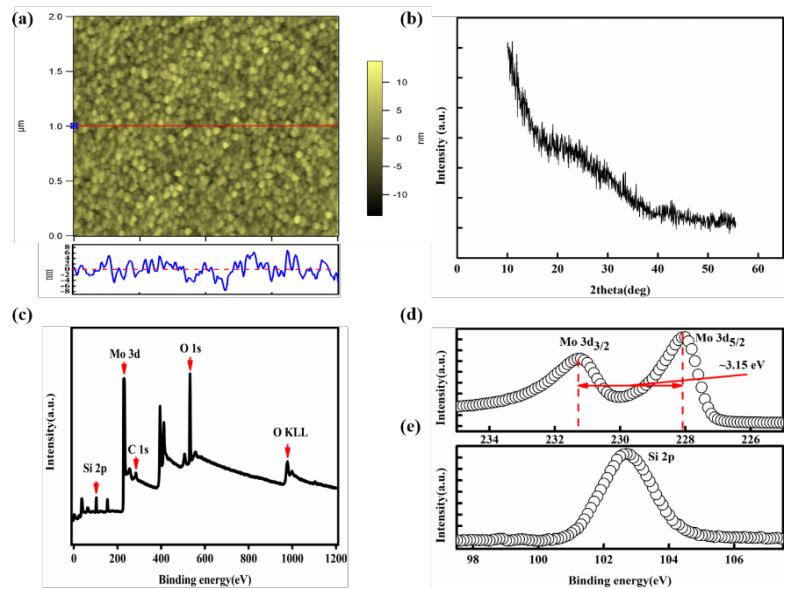

*Figure 1 Characterization of a-MoSi thin film,(a) AFM topographical image and the corresponding cross-section profile,(b) GIXRD pattern, (c) XPS full spectrum, (d) High-resolution Mo 3d XPS spectrum, (e) High-resolution Si 2p XPS spectrum.*

Figure 1 (a) shows the surface morphology and the corresponding cross-section profile of *a*-MoSi, demonstrating the uniform crack-free microstructure with root-mean-square surface roughness of 4.2 ± 0.1 nm. No typical diffraction peaks were observed in the GIXRD pattern of *a*-MoSi thin film, as shown in Figure 1 (b), evidencing the amorphous phase of this sample. Figure 1 (c) depicts the XPS full spectrum of *a*-MoSi thin film, indicating the existence of Mo and Si elements, as marked by the red arrow. One may note that the oxygen peaks also appear in the XPS spectrum, which is mainly attributed to surface adsorption. Additionally, high-resolution XPS results of Mo 3*d* and Si 2*p* are given in Figure 1 (d) and (e), respectively. Two distinct peaks centered at ~228 eV and ~231 eV, corresponding to $3d_{5/2}$ and $3d_{3/2}$, respectively, are observed in Mo 3*d* XPS spectrum (Figure 1(d)), due to the spin-orbital coupling effect [21]. The binding energies and the corresponding spin-orbit splitting energy (~3.15 eV) in *a*-MoSi thin film are consistent with the previously reported data in the case of $MoO_2$ [22], implying the molybdenum ions with the chemical state $Mo^{4+}$. As shown in Figure 1(e), the Si 2*p* XPS spectrum exhibits a high symmetric peak at ~103 eV, revealing the -4-valence state of silicon in the sample, as similarly observed in other Si-based amorphous materials [23].

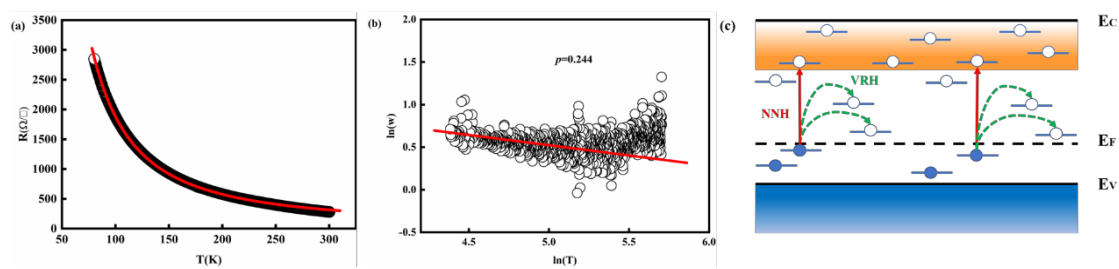

*Figure 2 (a) Temperature-dependent sheet resistance of a-MoSi thin film, (b) Zabrodskii analysis for temperature-dependent resistance data, (c) Schematic diagram for electronic transport. The*

*red solid lines in (a) and (b) show the fitting results.*

To determine the electronic transport mechanism, the temperature-dependence of sheet resistance was measured first, as shown in Figure 2 (a). The sample exhibits the typical semiconductor behavior in the measured temperature regime, *i.e.*, the resistance decreases with increasing the temperature. For a disordered solid system like amorphous material, variable-range hopping (VRH) model is generally utilized to analyze the transport mechanism, which describes the electron tunneling between localized sites [24, 25]. Given that all VRH models satisfy the modified Arrhenius expression, a self-consistent method, *i.e.*, Zabrodskii analysis [26], was used here to specifically decide the transport mechanism. The logarithmic derivative $w=-d(\ln(R))/d(\ln(T))=p*(T_0/T)^p$ is introduced, where $T_0$ is the characteristic parameter, $T$ is the temperature, and $p$ is the exponential term [27]. Different $p$ values correspond to different conduction mechanisms, for example, the thermal activation model ($p=1$), the Mott-VRH model ($p=1/4$) [28], and the Efros-Shklovskii (ES) VRH model ($p=1/2$) [29]. As shown in Figure 2 (b), a linear relationship is observed between $\ln(w)$ and $\ln(T)$, and the slopes of ~ 1/4 is determined, indicating the Mott-VRH conduction mechanism in *a*-MoSi thin film, as analogously reported in other amorphous binary compounds [19]. Meanwhile, the experimental data deviates from the theoretical model at high temperature, as predicted by the theory itself [28]. To further clarify the physical process in our sample, a concise schematic diagram is given in Figure 2 (c), wherein $E_C$, $E_F$ and $E_V$ correspond to the conduction band edge, Fermi energy level and valence band edge, respectively. As demonstrated above in GIXRD, *a*-MoSi thin film possesses the disorder microstructure, and thus the electronic wave-functions are naturally localized. Accordingly, the localized electrons below the Fermi level cannot directly enter the conduction band due to the lack of enough activated energy and the corresponding charge transport is achieved by the hopping of electrons between localized state sites. At higher temperature, *i.e.*, $k_BT$ beyond the average binding energy of the localized electron, phonon-assisted hopping allows access to the nearest neighbor hopping sites (distance is described by the horizontal scale) at a higher energy level, *i.e.*, the nearest neighbor hopping (NNH), as marked by the red solid arrows in Figure 2 (c). As the temperature decreases, the trade-off between the phonon-assisted term and the iso-energy tunneling probability, *i.e.*, between spatially proximal sites and sites which are spatially away but energetically accessible, leads to an optimization of hopping distance, which is described by the VRH of electrons between the localized states near the Fermi level, as marked by the green dotted arrow in Figure 2 (c).

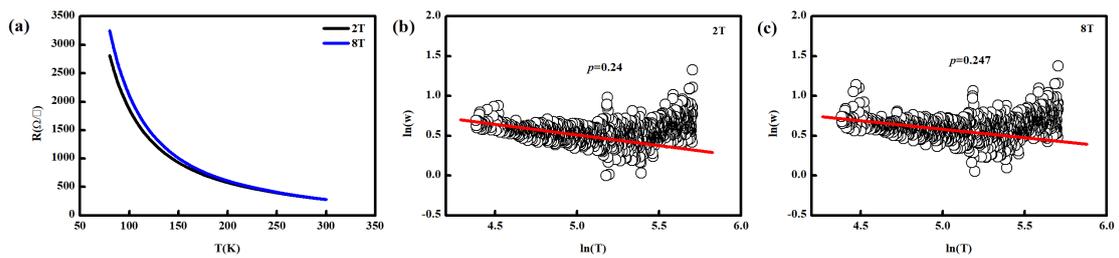

*Figure 3 (a) Temperature-dependent sheet resistance under different applied magnetic fields for a-MoSi thin film,(b) and (c) the corresponding Zabrodskii analysis for the raw data.*

Because the sample satisfies the Mott-VRH model rather well, the next step is to reveal the effect of the magnetic field on the localized state system. The temperature-dependent sheet resistance under different magnetic field was obtained, as shown in Figure 3 (a). Obviously, the sample shows the same semiconductor behaviors with a tiny discrepancy in the sheet resistance at different magnetic field, demonstrating that the Mott-VRH model is still valid under magnetic field, as shown in Figure 3 (b) and (c). Accordingly, the characteristic temperature $T_0$, calculated from the fitting curves, is slightly different with the magnetic field but does not show a consistent trend, which are $4.17\times10^5$ K (0T), $3.88\times10^5$ K (2T) and $4.60\times10^5$ K (8T). Despite the discrepancy caused by the experimental/fitting process, we here assume that the localization length $a_0$ does not change with the magnetic field at low magnetic field [30] as a prerequisite for the theoretical models discussed later.

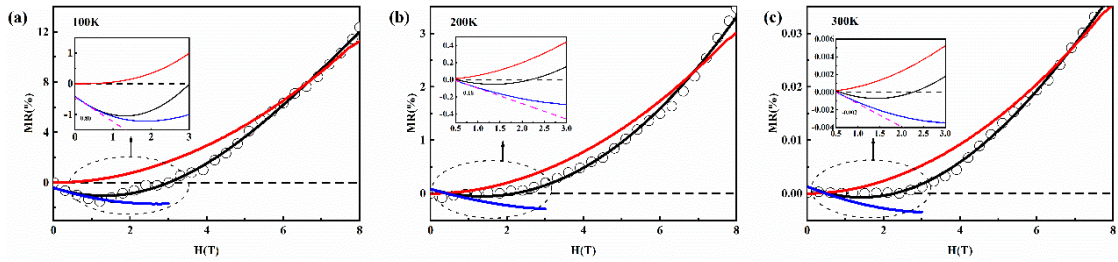

Figure 4 Magnetic-field dependence of MR for a-MoSi thin film at different temperatures, (a) 100 K, (b) 200 K, (c) 300 K. Circles are experimental data and solid lines are the fitting results. The inset graphs are the local decomposition corresponding to the magnetic-field regime marked by dash ellipses.

Since the magnetic field does not have a fundamental effect on the Mott VRH, we further focus on the behavior of the resistance changing continuously with the magnetic field. Figure 4 (a-c) presents the magnetic-field dependent MR at different temperatures. Here, MR is defined as MR=($R$(H)-$R$(0))/$R$(0)×100%, where $R$(H) and $R$(0) are the sheet resistance of a-MoSi thin film at magnetic field of $H$ and zero, respectively. With increasing magnetic field, the crossover from negative to positive

MR is observed. The negative MR characteristic, however, is weakened and disappeared at high temperature, *i.e.*, 300K. Similar phenomena were observed in some amorphous materials, for example, Ni$_x$Si$_{1-x}$ [19]. For the hopping conduction regime in disorder materials, the probability of hopping at the two localized sites can be easily varied under an applied magnetic field, leading to the MR.

To better understand the crossover from negative to positive MR, it is reasonably presumed that MR is the result of competition of positive MR (pMR) and negative MR (nMR):

$$MR = pMR + nMR \quad (1)$$

For Mott VRH case, pMR is accepted to be related to the wave-function shrinkage effect [30, 31]. Under the strong magnetic field, the wave-function will be squeezed, and the exponential tail of the wave-function will be less overlapped, resulting in the decreased hopping probability between two sites and thus the increased resistance, *i.e.*, positive MR. And the relationship between pMR and magnetic field $H$ can be expressed as [30, 31]

$$pMR = [t_2(T_0/T)^{1/4}/B_C^2] * H^2 \quad (2)$$

where $t_2 = (5/2016) \times 36 \approx 0.0893$ is a numerical constant and $B_C$ is the characteristic field.

On the other hand, nMR is accepted to be associated with the electron wave-function interference effect [32]. At low-field limit, the interference of electron wave-function must be enhanced to achieve the charge transition, leading to a nMR which is linear in magnetic field [30, 33]. Normally, the nMR caused by the interference enhancement at the low field can be expressed as [33]:

$$nMR = -(C_{sat}/B_{sat})*H \quad (3)$$

where $C_{sat}$ is the saturation constant and $B_{sat}=0.7(8/3)^{3/2}(1/a_0^2)(h/e)(T/T_0)^{3/8}$ is the effective saturation magnetic field.

Substituting Equation (2) and (3) into Equation (1), one can get

$$MR = [t_2(T_0/T)^{1/4}/B_C^2]*H^2 - (C_{sat}/B_{sat})*H \quad (4)$$

As shown in Figure 4, the experimental data are fitted well by equation (4), as marked by black curves, wherein the red curves represent the pMR part (the parabolic component) and the blue curves represent the nMR part (the linear component part at lower magnetic-field marked by magenta dashed lines). Evidently, the MR at the high field can be described rather well through the parabolic component, which is corresponding to the wave-function shrinkage caused by the magnetic field. However, at the low field, the combination of the wave function shrinkage effect and the electron wave-function interference enhancement requires both the linear term and the parabolic term, as shown in the insert graphs. Notably, this interference enhancement is more pronounced at lower temperature [31,33,34], which is consistent with the observed significant nMR at lower temperatures in our case (Figure 4 (a)). Additionally, the remaining nMR observed at high magnetic field (>3T) is mainly because the wave function shrinkage effect is not strong enough to completely overwhelm the interference enhancement and will not be discussed in detail here.

Furthermore, by deriving the Equation (4), the minimum value of MR can be expressed as

$$(MR)_{min} = -[B_C^2/4*t_2*(T_0/T)^{1/4}]*(C_{sat}/B_{sat})^2 \quad (5)$$

Obviously, the minimum value of MR is expected to be linearly related to the square of $C_{sat}/B_{sat}$, which could be determined through the slope of the low field nMR.

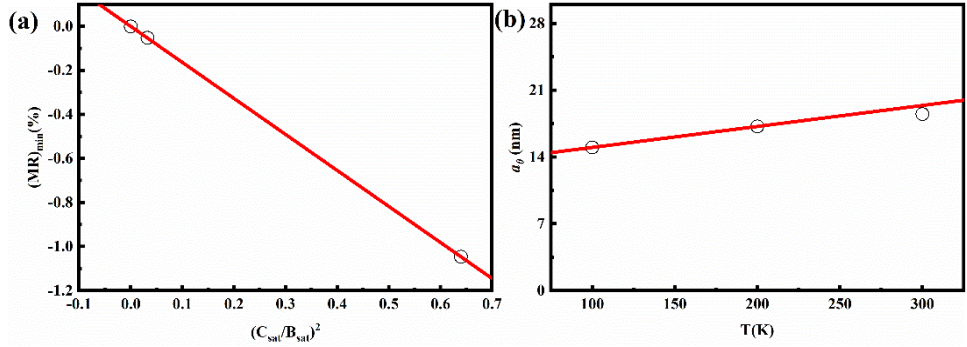

Figure 5 (a) The variation of the minimum value of MR, $(MR)_{min}$, as a function of $(C_{sat}/B_{sat})^2$ and (b) Temperature-dependent localization length $a_0$.

As shown in Figure 5 (a), the minimum value of MR shows a rather good linearity with $(C_{sat}/B_{sat})^2$, which is consistent with the theoretical prediction [33]. Consequently, the characteristic field, $B_C=6\hbar/[ea_0^2(T_0/T)^{1/4}]$, could be obtained from the slope and then the localization length $a_0$ could be determined as well. Figure 5 (b) shows the localization length at various temperatures for *a*-MoSi thin film. The increased localization length with temperature indicates the thermal excitation of the electron from the localized state to a more delocalized state, *i.e.*, quasi-extended state, as similarly reported in previous studies about the disorder-induced exciton localization [34, 35]. As the temperature increases, the interference effect is weakened due to the enhanced nearest neighbor hopping, resulting in the reduced possible hopping paths between two localized sites. Subsequently, the contribution of nMR is reduced and almost overwhelmed by pMR at high temperature (Figure 4(c)).

**Conclusions**

In summary, *a*-MoSi thin films were grown on silicon substrates at room temperature by RF magnetron sputtering technique, and the charge transport behaviors were studied by measuring the temperature-dependent resistance and magneto-resistance data. The temperature-dependent resistance was analyzed through a self-consistent method, and the Mott-VRH transport mechanism causing by the disorder-induced electronic wave-function localization was confirmed. Meanwhile, the crossover from negative to positive MR was observed in *a*-MoSi thin film with increasing magnetic field. The positive MR was primarily ascribed to the magnetic-induced electron wave-function shrinkage, whereas the negative MR was mainly related to the electron wave-function interference enhancement under magnetic field.

**Acknowledgements**

This work was supported by the National Key Research and Development Program of China (2017YFB0503300) and the Fundamental Research Funds for the Central Universities (Nos. 310201911cx024, 310201911fz048).